\documentclass[10pt]{article}
\usepackage[OE]{express}
\usepackage{color}
\usepackage{float}
\usepackage{afterpage}              
\usepackage{amssymb}
\usepackage{amsmath}
\usepackage{bm}
\usepackage{braket}
\usepackage{url}

\begin{document}
\title{Improved measurement of two-mode quantum correlations using a phase-sensitive amplifier}

\author{Tian Li,$^{1,*}$ Brian E. Anderson,$^1$ Travis Horrom,$^1$ Bonnie L. Schmittberger,$^1$ Kevin M. Jones,$^2$ and Paul D. Lett$^{1,3}$}

\address{$^1$Joint Quantum Institute, National Institute of Standards and Technology and the University of Maryland, College Park, MD 20742 USA\\
$^2$Department of Physics, Williams College, Williamstown, Massachusetts 01267 USA\\
$^3$Quantum Measurement Division, National Institute of Standards and Technology, Gaithersburg, MD 20899 USA}

\email{$^*$litian@umd.edu} 


\begin{abstract*}
We demonstrate the ability of a phase-sensitive amplifier (PSA) to pre-amplify a selected quadrature of one mode of a two-mode squeezed state in order to improve the measurement of  two-mode quantum correlations that exist before degradation due to optical and detection losses. We use four-wave mixing (4WM) in $^{85}$Rb vapor to generate bright beams in a two-mode squeezed state. One of these two modes then passes through a second 4WM interaction in a PSA configuration to noiselessly pre-amplify the desired quadrature of the mode before loss is intentionally introduced. We demonstrate an enhancement in the measured degree of intensity correlation and intensity-difference squeezing between the two modes. 
\end{abstract*}

\ocis{(190.4380) Nonlinear optics, four-wave mixing; (190.4970) Parametric oscillators and amplifiers; (270.6570) Squeezed states.} 

\bibliographystyle{unsrt}


\section{Introduction}

Nonclassical states of light have a wide range of applications in precision measurements, quantum imaging, optical communications, and quantum information science~\cite{QOfocus}. A severe limitation in using these quantum states is their sensitivity to loss, because loss adds noise. Specifically, quantum properties of two-mode states, such as continuous variable entanglement or two-mode squeezing, are quickly degraded if one or both modes are subject to loss. 

When significant downstream losses are present in a classical communication channel, classical amplifiers can be used before the loss to improve reception. An amplifier cannot increase the signal-to-noise ratio (SNR) of a communication channel, but it can increase the signal strength so that electronic noise in a receiver does not overwhelm the signal. The same concept can be applied to quantum communications. By adding the right kind of quantum-limited amplifier before downstream loss, selected properties of a quantum state can be maintained even in the presence of loss.

The most common type of amplifier is a phase-insensitive amplifier (PIA) which amplifies both quadratures of the input channel. A PIA necessarily has an open input port which admits vacuum noise, and in the limit of large PIA gain,  the signal will suffer an SNR degradation of 3~dB.  In contrast, a phase-sensitive amplifier (PSA) has no open ports and can noiselessly amplify one quadrature of the input. Provided the signal is encoded in the appropriate quadrature, the SNR remains unchanged~\cite{Caves1982}.

One form of downstream optical loss is the imperfect quantum efficiency of a photodetector. The concept of using optical amplification to compensate for imperfect detection efficiency has been considered previously. It has been shown theoretically that by pre-amplifying the observed quadrature component of the input electric field, one can compensate for the non-unity detector quantum efficiency in a homodyne detector~\cite{Leonhardt1994, Marchiolli1998, Kim1997, Ahmad2000, Fossier2009}. Experimentally,  noiseless optical amplification has been used in the context of quantum non-demolition measurements~\cite{Poizat1993, Bencheikh1995, Bencheikh1996, Grangier1998, Levenson1993, Levenson1993a, Levenson1999}. In these experiments, noiseless amplification was used to overcome downstream propagation losses and non-unity detector quantum efficiency, although in each case the input signal before amplification was classical. Lam $et$ $al.$~\cite{Lam1997} used an electro-optic feed forward scheme to produce an output state with an SNR close to that of the input state, outperforming a PIA. An amplitude squeezed state was used as the input to the device.  Ulanov $et$ $al.$~\cite{Ulanov2015} demonstrated using noiseless optical amplification to restore entanglement in the presence of loss in a probabilistic manner in the photon-counting regime, i.e., entanglement distillation. \textcolor{black}{Alon $et$ $al.$~\cite{Alon} used a PSA to demonstrate phase-sensitive amplification and deamplification of a non-classical, quadrature-squeezed state.}

In this paper we report the use of a quantum-limited optical PSA to pre-amplify one mode of a nonclassical two-mode state of light in order to overcome downstream optical losses for that mode.  
\textcolor{black}{We make a systematic investigation of using a PSA to improve the measurement of  correlations and squeezing levels that were present in the input state of light, in particular compensating for various levels of non-unity detector quantum efficiency.}  \textcolor{black}{While Alon $et$ $al.$~\cite{Alon} demonstrate a proof-of-principle experiment, here we investigate more fully the behavior with respect to varying loss in the system.} 

\begin{figure}
\begin{centering}
\includegraphics[width=5.5in]{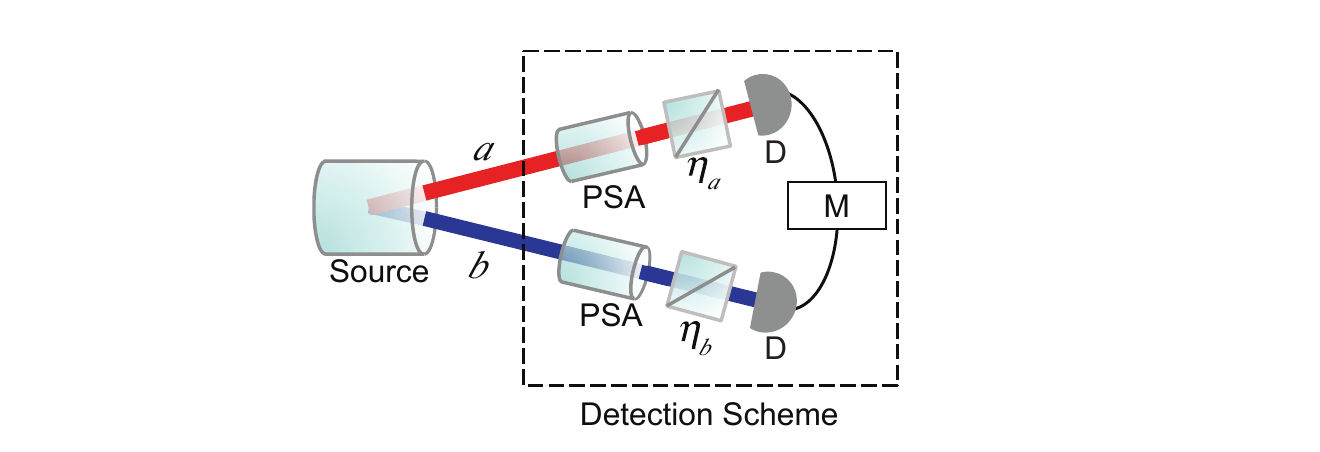}
\end{centering}
\caption{Conceptual schematic of the experiment. The source produces two modes $a$ and $b$ that are quantum correlated. The non-unity quantum efficiency of the detection and any other optical losses are symbolized by two beamsplitters with transmission $\eta_a$ and $\eta_b$. D denotes ideal detectors with perfect quantum efficiency. M represents the processing of the detected signals to produce information about the quantum state in modes $a$ and $b$.}
\label{fig_conceptual_diagram}
\end{figure}

Figure~\ref{fig_conceptual_diagram} shows the conceptual basis of the experiment. The source produces a quantum state with modes $a$ and $b$ whose properties we wish to measure. Measurements made by the detectors are contaminated by vacuum noise coupled in by optical losses ($\eta_a$ and $\eta_b$). These losses include less-than-perfect detector efficiencies. Inserting a PSA in each arm, adjusted to amplify the desired quadrature, allows one to reduce the influence of the vacuum noise coupled in by the optical losses and thus make a better measurement of the quantum properties of the state of modes $a$ and $b$.  Depending on the measurement, the processing of the data may include electronic gain adjustments to compensate for the optical gain introduced by the PSAs. 

For technical simplicity in our proof-of-principle experiment we use a single PSA and mitigate the effect of losses in only one arm of the apparatus. The detectors $D$ are intensity detectors and the PSA is adjusted to amplify the intensity quadrature of the light in one arm. The quantum properties of the modes $a$ and $b$ which we choose to measure are the correlation coefficient of the optical intensities and the squeezing of the intensity difference between these two modes. To explore the effect of the PSA on these measurements we intentionally introduce a known loss after the PSA and record the effect that this loss has on the measured intensity correlation coefficient \textcolor{black}{and twin-beam noise reduction, or intensity-difference squeezing, between the modes.}

\section{Experimental setup}

\begin{figure}
\begin{centering}
\includegraphics[width=5in]{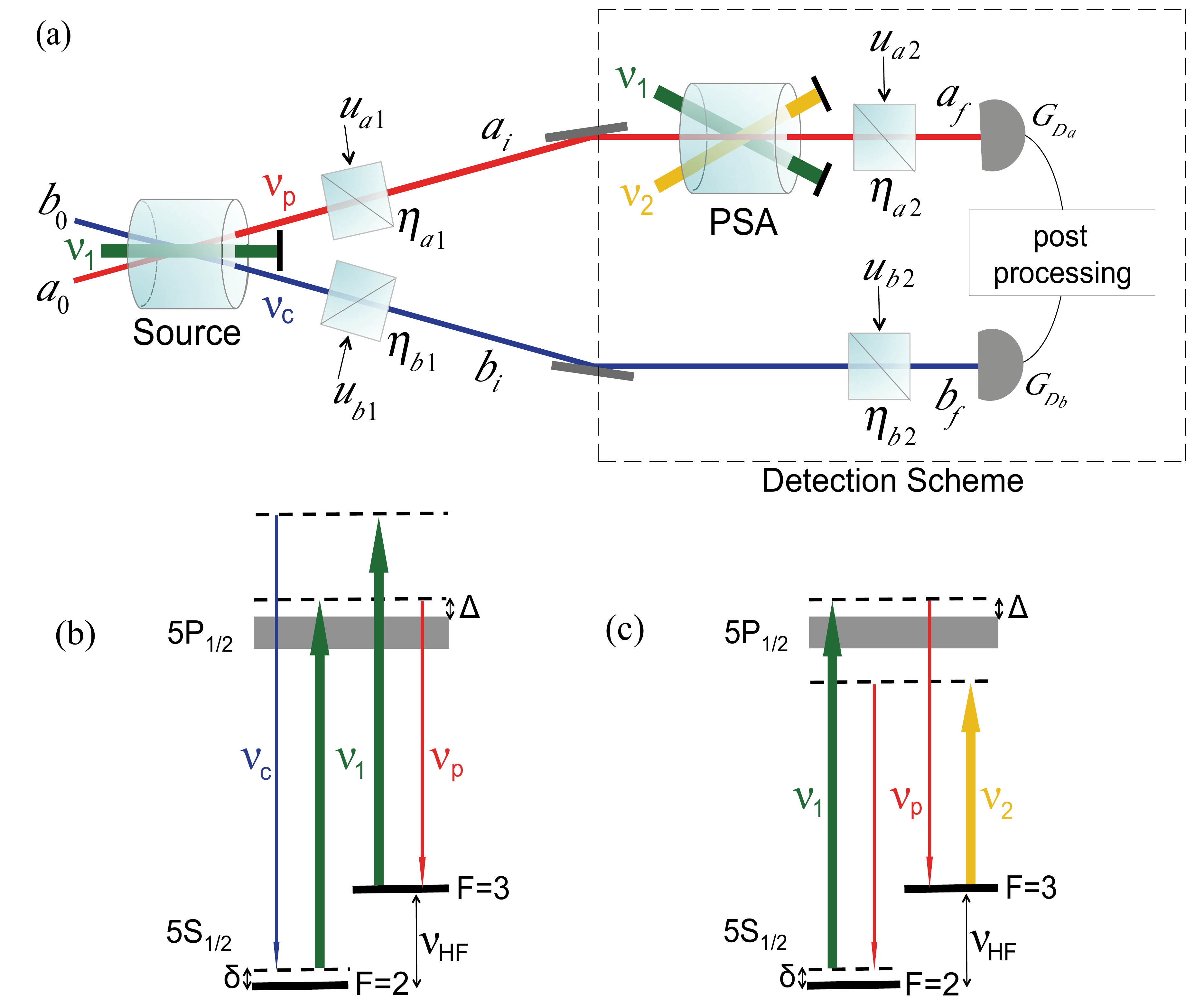}
\end{centering}
\caption{Experimental setup and 4WM schemes showing atomic energy levels  in $^{85}$Rb and laser tunings. (a) Experimental setup. The source generates a two-mode squeezed state. $\eta$'s are the transmissions of beamsplitters that represent losses: $\eta_{\text{a1}}$ is the probe transmission before the PSA and $\eta_{\text{a2}}$ is the probe transmission after the PSA representing all the downstream losses including imperfect detector efficiency. We vary the value of $\eta_{\text{a2}}$ by intentionally introducing extra loss using a half-wave plate and a polarizing beamsplitter. The transmission $\eta_{\text{b2}}$ includes the effect of imperfect detector efficiency on the measurement of the conjugate beam. $G_\text{Da}$ and $G_\text{Db}$ are the gains of the probe and conjugate detectors, respectively. (b) 4WM scheme in the source cell. $\nu_p$, $\nu_c$ and $\nu_1$ are the optical frequencies of probe, conjugate and pump beams, respectively, \textcolor{black}{and $\nu_p + \nu_c = 2 \nu_1$}. (c) 4WM scheme in the PSA cell. $\nu_1$, $\nu_2$ are the optical frequencies of the two pump beams, and $\nu_p$ is the optical frequency of the probe beam, \textcolor{black}{and $\nu_1 + \nu_2 = 2 \nu_p$}. For both (b) and (c): the width of the excited state in the level diagram represents the Doppler broadened line, $\Delta$ is the one-photon detuning, $\delta=-4$~MHz is the two-photon detuning, and $\nu_{HF}=3.036$~GHz is the hyperfine splitting in the electronic ground state of $^{85}$Rb.}
\label{fig_set_up}
\end{figure}

A detailed description of our experiment is shown in Fig.~\ref{fig_set_up}(a).  Both the source and the PSA are implemented using four-wave mixing (4WM) in $^{85}$Rb vapor. Details of the source are discussed in~\cite{McCormick2007} and of the PSA in~\cite{Corzo2011}. In both cases, the $^{85}$Rb atoms are contained in a 12~mm glass cell.  The source and PSA cells are heated to 112~$^{\circ}$C and 86~$^{\circ}$C, respectively. Figure~\ref{fig_set_up}(b) shows the atomic energy levels and detunings used in the 4WM process of the source.  We seed mode $a_0$ with a weak coherent beam (0.1~mW, 300~$\mu$m $1/e^2$ diameter) tuned to the probe frequency $\nu_p$ and let mode $b_0$ be the vacuum. The source is driven by a strong pump (350~mW, 800~$\mu$m $1/e^2$ diameter) that can be regarded as classical. The output modes $a_i$ (at the probe frequency $\nu_p$) and $b_i$ (at the conjugate frequency $\nu_c$), are quantum correlated~\cite{McCormick2007}. Losses are modeled by beamsplitters with transmission $\eta_{\text{a1}}$ and $\eta_{\text{b1}}$. These represent both losses inside the source cell as well as external losses. The calculation of loss (as well as gain) inside the source cell is discussed in the Appendix. In the absence of the PSA and any intentionally introduced losses, we measure -5.8~dB of intensity-difference squeezing, which is roughly constant over an analysis frequency range of 0.5~MHz to 2.5~MHz.

Probe mode $a_i$ passes through a second $^{85}$Rb cell which is pumped by two strong classical pumps, one at $\nu_1$ (the same as the source pump) and the other one at $\nu_2$, approximately 6~GHz downshifted relative to $\nu_1$, as shown in Fig.~\ref{fig_set_up}(c). The two pump beams (100~mW each, 500~$\mu$m $1/e^2$ diameter) and the probe beam are in a plane with an angular separation of 0.6 degrees between the pumps. The PSA implemented in this work is the same as in~\cite{Li2016, Corzo2012, Corzo2011}.




The probe field created in the source is amplified or deamplified in the PSA depending on the PSA phase

\begin{equation}
\phi_{\text{PSA}}=(\phi_{1} - \phi_p) + (\phi_{2} - \phi_p),
\label{eq:phase}
\end{equation}
where $\phi_{1}$ and $\phi_{2}$ are the optical phases of the pump beams and $\phi_p$ is the probe optical phase. Assuming that the input signal is encoded in a single quadrature, there are two choices of PSA phase for which the PSA will noiselessly amplify this input quadrature. For one choice, this quadrature intensity experiences a gain $G_{\text{PSA}}$ and for the other, it experiences a deamplification $1/G_{\text{PSA}}$. In our experiment we keep $\phi_{\text{PSA}}$ such that the input intensity of the probe beam always sees gain $G_{\text{PSA}}$. Since $G_{\text{PSA}}$ is a function of pump power, one photon detuning $\Delta$ (see Fig.~\ref{fig_set_up}(c)), and the number density of $^{85}$Rb atoms in the cell, adjusting these parameters allows us to vary $G_{\text{PSA}}$. In the experiment, we change the one photon detuning $\Delta$ to obtain different values for $G_{\text{PSA}}$.

All the downstream losses experienced by the probe after the PSA are modeled by a beamsplitter as shown in Fig.~\ref{fig_set_up}(a). Its transmission $\eta_{\text{a2}}$ is determined by the transmission of the downstream optics including the exit window of the PSA cell, the quantum efficiency of the detector photodiode and an intentionally introduced extra loss from the combination of a half-wave plate and a polarizing beamsplitter. An additional contribution to $\eta_{a2}$ comes from the non-zero electronic noise floor of the detection electronics. The finite separation of the optical noise power from the detector electronic noise floor adds noise equivalent to that of an attenuator whose transmission is $1-{10}^{-\frac{\text{S}}{10}}$, where $\text{S}$ is the noise power separation in dB~\cite{Lvovsky2007}.  Losses experienced by the conjugate are modeled by a beamspliter with transmission $\eta_{b2}$. This includes the less-than-perfect quantum efficiency of the detector photodiode and the finite separation of the conjugate optical noise power from the detector electronic noise floor. 
 
The AC and DC components of the intensities of the two modes $a_f$ and $b_f$ are recorded separately. We post-process the AC time traces by filtering them between 0.5~MHz to 2.5~MHz using a 4th-order Butterworth band-pass filter with a slope of 80~dB/decade on both the low and high pass edges. The resulting time traces are used to determine the intensity correlation coefficient and the intensity-difference squeezing between the two beams by employing Eqs.~(\ref{eq_xc}) and~(\ref{eq_sq}) derived in the following section.

\section{Model}

We use a simple quantum-mechanical model to simulate the experiment. We label the optical field operators for the modes shown in Fig.~\ref{fig_set_up}(a) as $\hat{a}_{0,i,f}$, $\hat{b}_{0,i,f}$, and $\hat{u}_{a1,a2,b1,b2}$. Since there is no PSA between the two beamsplitters in the conjugate path, we can combine the two beamsplitters into one with transmission $\eta_b = \eta_{b1} \cdot \eta_{b2}$ and operator $\hat{u}_{b}$ associated with the vacuum field coupling into it. We treat all pump beams classically. The vectors $\hat{\vec{{V}_0}}$ and $\hat{\vec{{V}_f}}$ are the input and output field operators defined by

\begin{equation}
\hat{\vec{{V}_0}}=
\begin{pmatrix}
\hat{a}_0\\ \hat{a}^\dagger_0\\ \hat{b}_0\\ \hat{b}^\dagger_0
\end{pmatrix} \hspace{4mm} \text{and} \indent 
\hat{\vec{{V}_f}}=
\begin{pmatrix}
\hat{a}_f\\ \hat{a}^\dagger_f\\ \hat{b}_f\\ \hat{b}^\dagger_f
\end{pmatrix}.
\end{equation}
The source and PSA can then be described by the matrices 

\begin{equation}
\bm{F_1}=
\begin{pmatrix}
\cosh{r} &0 &0 &\sinh{r}\\ 0 &\cosh{r} &\sinh{r} &0\\ 0 &\sinh{r} &\cosh{r} &0\\ \sinh{r} &0 &0 &\cosh{r}
\end{pmatrix} 
\end{equation}
 and
\begin{equation}
\bm{F_2}=
\begin{pmatrix}
\cosh{s} &e^{i\phi_{\text{PSA}}}\sinh{s} &0 &0\\ e^{-i\phi_{\text{PSA}}}\sinh{s} &\cosh{s} &0 &0\\ 0 &0 &1 &0\\ 0 &0 &0 &1 
\end{pmatrix}, 
\end{equation}
respectively. Here, $r$ and $s$ are the squeezing parameters which are related to the gains of the source and PSA via $G_{\text{source}} = \cosh^2 r$ and $G_{\text{PSA}} = e^{2 s}$. $\phi_{\text{PSA}}$ is the phase of the PSA defined in Eq.~(\ref{eq:phase}). We keep $\phi_{\text{PSA}} = 0$ such that the intensity gain of the probe beam is always $G_{\text{PSA}}$.

The experiment can then be described by the transformation of field operators 

\begin{equation}
\hat{\vec{{V}_f}}=\bm{T_2}\left( \bm{F_2} \left[ \bm{T_1} (\bm{F_1} \hat{\vec{{V}_0}}) + \hat{\vec{{L}_1}}\right]\right) +\hat{\vec{{L}_2}},
\label{eq:physics}
\end{equation}
where the matrices $\bm{T_1}$ and $\bm{T_2}$ describe the transmission of the beamsplitters, and vectors $\hat{\vec{{L}_1}}$ and $\hat{\vec{{L}_2}}$ contain the field operators for the vacuum modes coupled in by optical losses:

\begin{equation}
\bm{T_1}=
\begin{pmatrix}
\sqrt{\eta_{\text{a1}}}& 0& 0& 0\\ 0& \sqrt{\eta_{\text{a1}}}& 0& 0\\ 0& 0& \sqrt{\eta_{\text{b}}}& 0 \\ 0& 0& 0& \sqrt{\eta_{\text{b}}} 
\end{pmatrix}, 
\end{equation}

\begin{equation}
\bm{T_2}=
\begin{pmatrix}
\sqrt{\eta_{\text{a2}}}& 0& 0& 0\\ 0& \sqrt{\eta_{\text{a2}}}& 0& 0\\ 0& 0& 1& 0 \\ 0& 0& 0& 1 
\end{pmatrix},
\end{equation}

\begin{equation}
\hat{\vec{{L}_1}}=
\begin{pmatrix}
i\sqrt{1-\eta_{\text{a1}}}\hat{u}_{\text{a1}}\\ -i\sqrt{1-\eta_{\text{a1}}}\hat{u}^\dagger_{\text{a1}} \\ i\sqrt{1-\eta_{\text{b}}}\hat{u}_{\text{b}}\\ -i\sqrt{1-\eta_{\text{b}}}\hat{u}^\dagger_{\text{b}}
\end{pmatrix},
\end{equation}

\begin{equation}
\hat{\vec{{L}_2}}=
\begin{pmatrix}
i\sqrt{1-\eta_{\text{a2}}}\hat{u}_{\text{a2}}\\ -i\sqrt{1-\eta_{\text{a2}}}\hat{u}^\dagger_{\text{a2}}\\ 0 \\ 0 
\end{pmatrix}.
\end{equation}
From Eq.~(\ref{eq:physics}), we can derive the field operators $\hat{a}_f$ and $\hat{b}_f$ and the number operators $\hat{n}_a = \hat{a}^\dagger_f \hat{a}_f$ and $\hat{n}_b = \hat{b}^\dagger_f \hat{b}_f$ for the output modes. $\hat{n}_a$ and $\hat{n}_b$ allow us to calculate the expectation values of quantities such as the intensity correlation coefficient and the intensity-difference squeezing. 

The intensity correlation coefficient is calculated from

\begin{equation}
M_{\text{XC}} = \frac {\braket{(\hat{n}_a - \braket{\hat{n}_a})(\hat{n}_b - \braket{\hat{n}_b})}}{\sqrt{\Delta^2 \hat{n}_a}\sqrt{\Delta^2 \hat{n}_b}},
\label{eq_xc}
\end{equation}
where $\Delta^2 \hat{n}_a$ and $\Delta^2 \hat{n}_b$ are the variances of the intensities. Since we are only looking at AC components of the signal, $\braket{\hat{n}_a}=\braket{\hat{n}_b}=0$.

The intensity-difference squeezing is calculated from

\begin{equation}
M_{\text{SQ}} = -10 \log_{10} \left[\frac{\Delta^2 (\hat{n}_a - \hat{n}_b)}{\Delta^2 \hat{n}_{\text{SN}}} \right],
\label{eq_sq}
\end{equation}
where $\Delta^2 \hat{n}_{\text{SN}}$ is the shot noise, which is defined as the variance of the intensity difference of two coherent beams having the same intensities as the measured probe and conjugate beams.

\textcolor{black}{To compare the predictions of this model to measurements requires values of the various gain and loss parameters. The Appendix describes how to extract source parameters from auxiliary measurements of beam intensities. The PSA gain and the inserted losses are measured directly. No other fitting parameters are required to generate the theoretical curves shown in the following section. }

\section{Results}

\begin{figure}[t]
  \centering
  \includegraphics[width=4.2in]{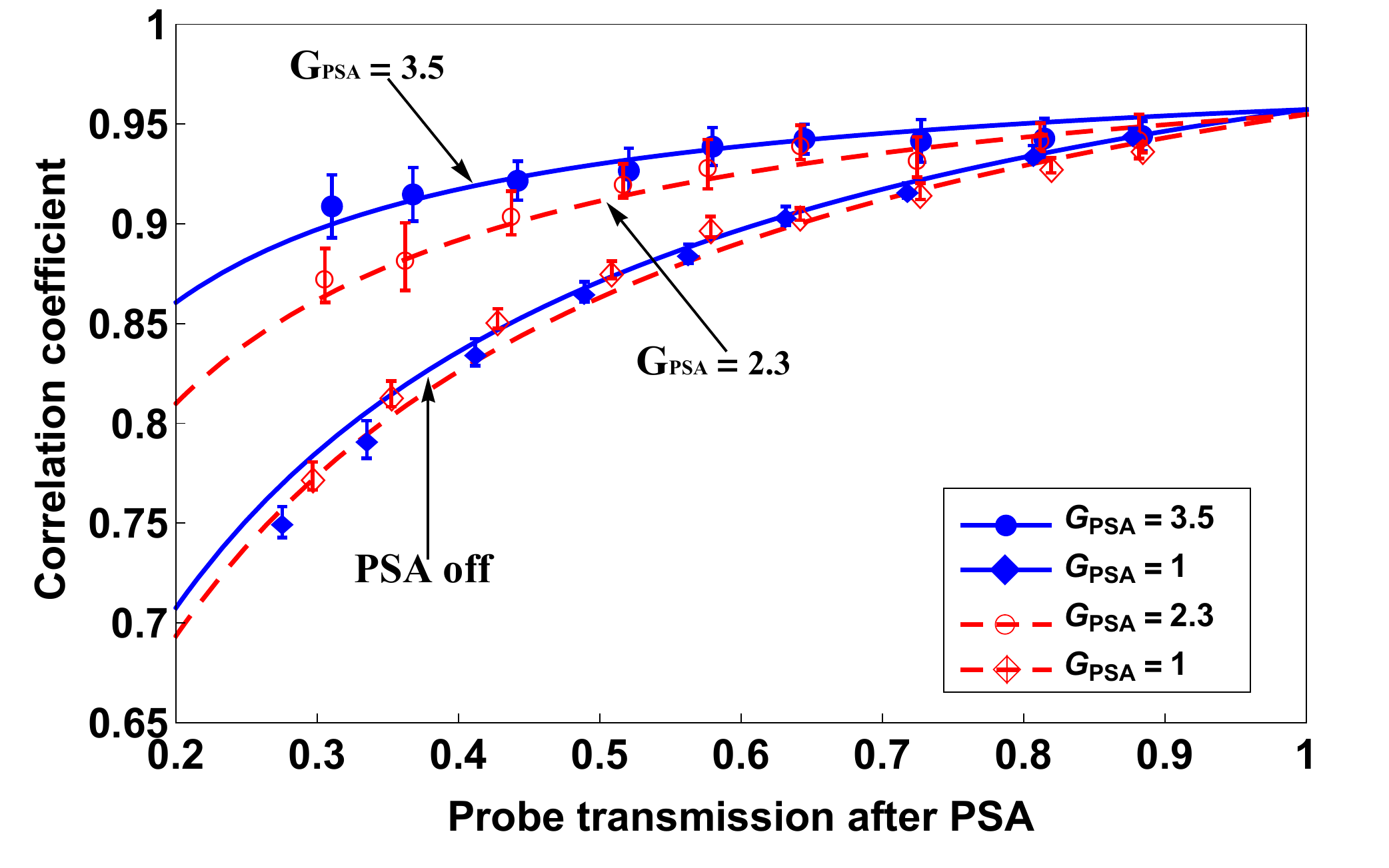}
  \caption
   {Intensity correlation coefficient of the probe and conjugate beams as a function of probe transmission $\eta_{\text{a2}}$. Blue points and lines are for the source gain $G_{\text{source}}$ = 3.3.  Red points and dashed lines are for the source gain $G_{\text{source}}$ = 3.0.  The diamonds (both open and solid) are for the PSA cell removed from the probe path (i.e., $G_{\text{PSA}} = 1$). The circles are for the PSA present with a gain of 2.3 (open circles) and 3.5 (solid circles), respectively. The solid and dashed lines are theoretical predictions calculated from Eqs.~(\ref{eq:physics}) and~(\ref{eq_xc}) using the source parameters given in the Appendix.}
\label{fig:XCorr}
\end{figure}

In Fig.~\ref{fig:XCorr}, we plot the measured intensity correlation coefficient of the probe and conjugate beams as a function of the probe transmission $\eta_{\text{a2}}$ after the PSA. To operate the PSA at different gains, we adjust the one photon detuning $\Delta$ (see Fig.~\ref{fig_set_up}(c)).  Since the two 4WM processes share a pump beam, changing the one photon detuning $\Delta$ changes the gain of the source, $G_\text{source}$, as well. In order to have a consistent preparation of the two-mode squeezed state, we adjust the pump power going to the source so that the intensity-difference squeezing produced by the source stays at -5.8~dB.  Specifically, for the measurements shown in Fig.~\ref{fig:XCorr} we choose $\Delta= 1.4$~GHz and 1.3~GHz to produce a PSA gain of 2.3 and 3.5, respectively. As we change the detuning, the losses in the source also change, requiring a different source gain ($G_\text{source}$ = 3.0 and 3.3, respectively) to maintain the squeezing level (see Appendix). 

Figure~\ref{fig:XCorr} shows that adding a PSA increases the measured intensity correlation coefficient. \textcolor{black}{The blue (red) circles show data taken with  $G_{\text{PSA}} = 3.5$ (2.3), and the diamonds show data taken when the PSA is removed (i.e.  $G_{\text{PSA}} = 1$).  $G_{\text{source}}$ for each set of data is given in the figure caption.} The vertical error bars on the data points are one standard deviation statistical uncertainties from 60 time traces. Data sets sharing the same $\Delta$ but with the PSA on and off do not share the same transmission values owing to the variation in the optical noise power separations from the detector electronic noise floor~\cite{Lvovsky2007}. The highest value of the transmission after the PSA is 0.88 (the rightmost open red diamond), which is determined by the optical transmission of 0.99, the detector photodiode quantum efficiency of 0.90, and the loss of 1.6~\% associated with a noise power separation of 18 dB from the detector electronic noise floor. 
Extrapolating to zero loss, the intensity correlation coefficients are 0.955 and 0.957 for $G_{\text{source}} = 3.0$ and 3.3, respectively. For an ideal source with no losses, the correlation coefficient is related to the squeezing parameter $r$ by $M_{XC}=\tanh 2r$. In the limit of large $r$ (high $G_\text{source}$),  $M_{XC}$ approaches one. Figure~\ref{fig:XCorr} clearly shows that the intensity correlation coefficient is improved when the PSA is present, and greater improvement is achieved with higher $G_{\text{PSA}}$, e.g., the data points with $G_{\text{PSA}}=3.5$ are closer to the correlation coefficient  that would be measured with an ideal detector than the ones with  $G_{\text{PSA}}=2.3$.

In Fig.~\ref{fig:RawSQZ} we plot the measured intensity-difference squeezing versus the probe transmission $\eta_{\text{a2}}$ after the PSA. The shot noise is measured from the time traces of two coherent beams that have the same intensities as the modes $a_f$ and $b_f$ for each transmission. The squeezing is obtained from the time traces of the measured probe and conjugate intensities according to Eq.~(\ref{eq_sq}). In Fig.~\ref{fig:RawSQZ}, when the PSA is absent ($G_{\text{PSA}} = 1$), we measure a best intensity-difference squeezing of -5.8 dB. As we lower the transmission, we gradually lose the squeezing as expected; at the lowest two transmissions, we lose the squeezing completely. For high transmission values, turning on the PSA destroys the squeezing. When the PSA is on, the probe beam power is amplified by the PSA, which causes a power imbalance between the probe and conjugate beams. Unlike the correlation coefficient, squeezing is affected by changing the relative probe and conjugate powers. As the probe transmission decreases, the power of the probe mode $a_f$ approaches that of the conjugate mode $b_f$, and squeezing is \textcolor{black}{partially} restored.

\begin{figure}[H]
  \centering
  \includegraphics[width=4.2in]{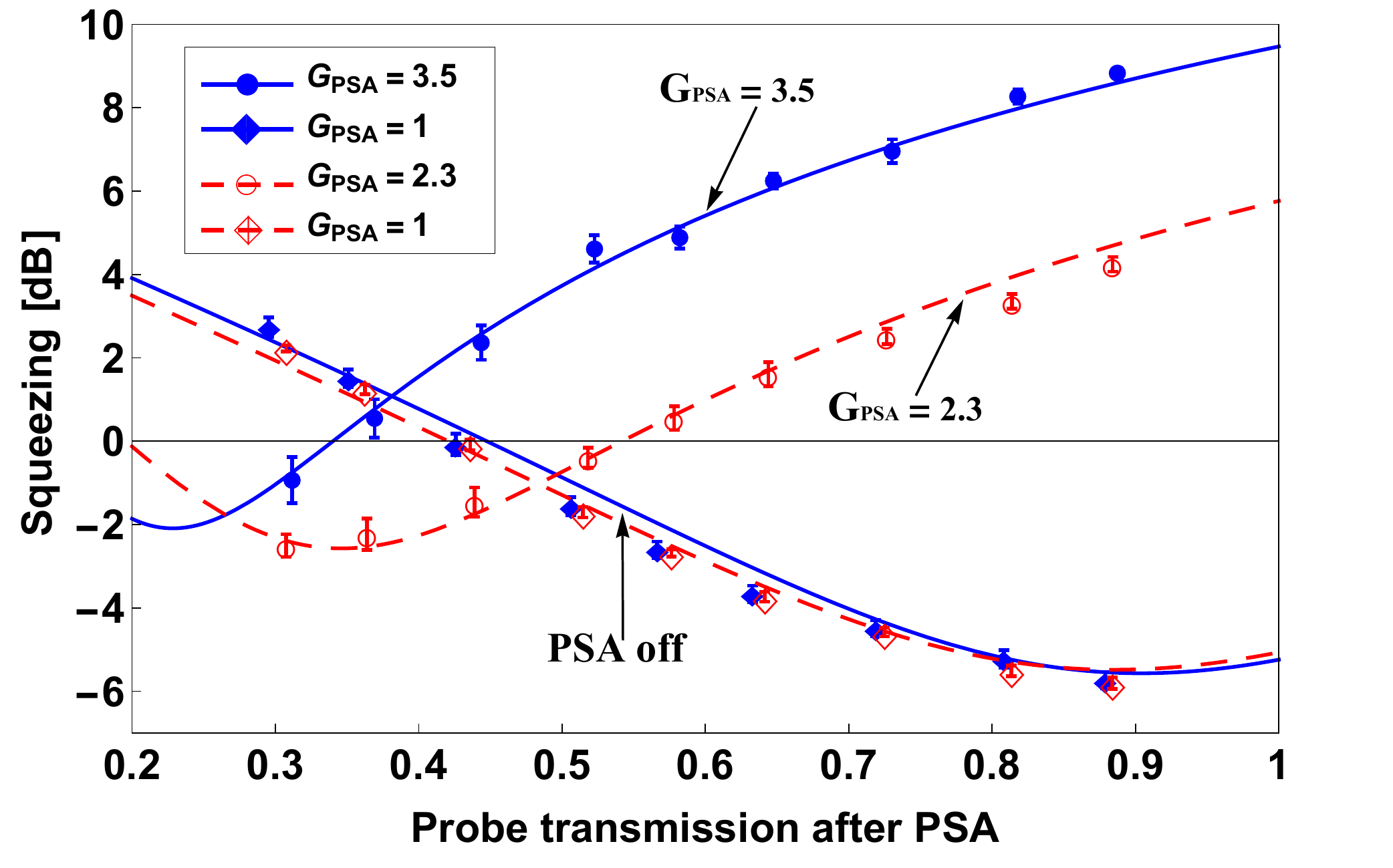}
  \caption
   {Intensity-difference squeezing measured from the time traces of the intensities of modes $a_f$ and $b_f$, as a function of probe transmission $\eta_{\text{a2}}$. Blue points and lines are for the source gain $G_{\text{source}}$ = 3.3.  Red points and dashed lines are for the source gain $G_{\text{source}}$ = 3.0.  The diamonds (both open and solid) are for the PSA cell removed from the probe path (i.e., $G_{\text{PSA}} = 1$). The circles are for the PSA present with a gain of 2.3 (open circles) and 3.5 (solid circles), respectively. The solid and dashed lines are theoretical predictions calculated from Eqs.~(\ref{eq:physics}) and~(\ref{eq_sq}) using the source parameters given in the Appendix.}
\label{fig:RawSQZ}
\end{figure}

The power imbalance between probe and conjugate caused by the PSA can be compensated by adjusting the gain $G_\text{Da}$ of the probe detector. In particular, by adjusting $G_\text{Da}$ so that $G_{\text{PSA}} \times \eta_{\text{a2}} \times G_{\text{Da}} = 1$, the probe power is the same as would be measured by an ideal detector in mode $a_i$. Similarly, we adjust $G_\text{Db}$ so that $ \eta_{\text{b2}} \times G_{\text{Db}} = 1$. Since we do not vary $\eta_{\text{b2}}$, $G_{\text{Db}}$ is a constant.
By performing this gain adjustment, the shot noise is determined by the intensity-difference noise of two \textcolor{black}{shot-noise-limited} coherent beams that have the same intensities as the two modes $a_i$ and $b_i$ measured by two ideal detectors, regardless of $G_{\text{PSA}}$ or $\eta_{\text{a2}}$. 

\begin{figure}[H]
  \centering
  \includegraphics[width=4.25in]{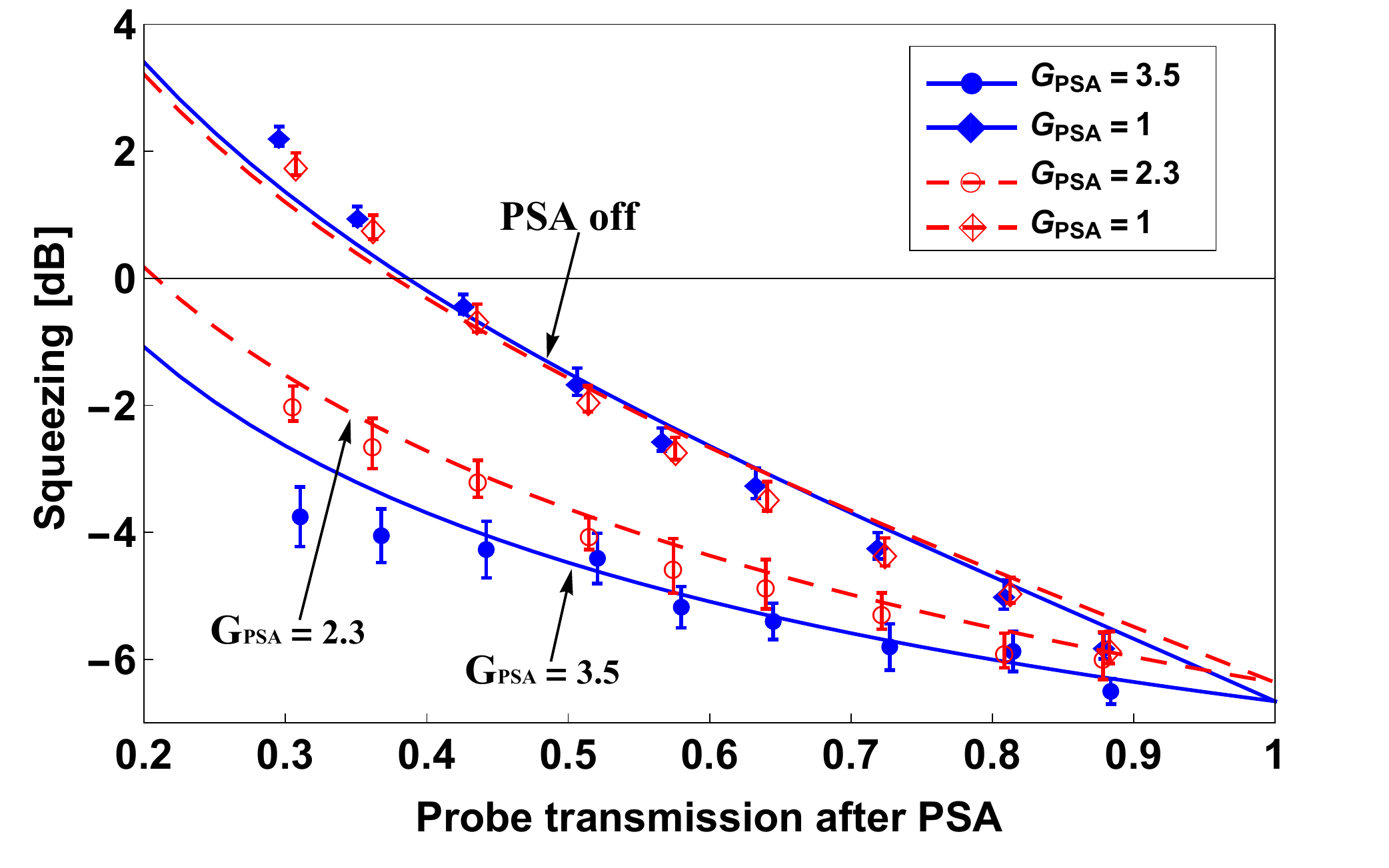}
  \caption
   {Intensity-difference squeezing measured with the detector gain adjustments described in the text. Blue points and lines are for the source gain $G_{\text{source}}$ = 3.3.  Red points and dashed lines are for the source gain $G_{\text{source}}$ = 3.0.  The diamonds (both open and solid) are for the PSA cell removed from the probe path (i.e., $G_{\text{PSA}} = 1$). The circles are for the PSA present with a gain of 2.3 (open circles) and 3.5 (solid circles), respectively. The solid and dashed lines are theoretical predictions calculated from Eqs.~(\ref{eq:physics}) and~(\ref{eq_sq}) using the source parameters given in the Appendix.}
\label{fig:RescaledSQZ}
\end{figure}

By applying the appropriate value of $G_\text{Da}$ to the measured time traces at each value of $\eta_{\text{a2}}$ and $G_{\text{PSA}}$ in Fig.~\ref{fig:RawSQZ}, we obtain the points in Fig~\ref{fig:RescaledSQZ}. With this gain adjustment, points which previously did not show squeezing now do. 
The larger the $G_\text{PSA}$ is, the better the improvement in the measured intensity-difference squeezing. For example, without the PSA the squeezing vanishes at approximately 60~\% loss, while when $G_{\text{PSA}}=3.5$, all of the measured points show squeezing and the theory shows that squeezing of -1~dB would still be measured down to even 80~\% loss. 

\textcolor{black}{The squeezing levels in Figs.~\ref{fig:RawSQZ} and~\ref{fig:RescaledSQZ} become positive at small transmission. This is related to the asymmetric treatment of the two modes in the experiment.  Both beams have thermal statistics coming from the source.  The additional attenuation is present only in one beam and slowly changes this beam's statistics to that of a vacuum coherent state at large attenuation. Thus, even with the detector gain adjustment, the probe beam statistics are different from those of the conjugate beam, which remains a thermal beam, and this gives rise to the anti-squeezing in the intensity-difference.}

%
%

\section{Conclusions}
We experimentally demonstrate the use of an optical PSA to reduce the effect of non-unity quantum efficiency of an intensity detector \textemdash~an important limitation in making continuous variable measurements of quantum states. Use of the PSA allows us to more accurately measure the intensity correlation coefficient and the intensity-difference squeezing of a two-mode squeezed state. For large PSA gain, the system of a PSA followed by a detector with imperfect quantum efficiency approaches a ``perfect detector~\cite{Bencheikh1995}.'' We compare experimental measurements taken with various PSA gains and levels of intentionally introduced loss to a simple quantum-mechanical model and find reasonable agreement. The type of PSA used here is capable of operating on multiple spatial modes~\cite{Corzo2012}, and thus may be useful for overcoming detector efficiency limitations in quantum imaging applications.

\section*{Appendix}
\section*{Calculation of gain and loss inside the source cell}

The correlation properties of a two-mode squeezed state emitted by a lossless source are solely determined by its gain. In the experiment, however, the atomic absorption and scattering from the $^{85}$Rb atoms in the source cell inevitably introduce loss. Therefore, a complete description of the source requires a distributed gain and loss model~\cite{PhysRevA.78.043816, Jasperse:11}. Here we adopt a simpler model that assumes an ideal lossless source followed by loss as shown in Fig.~\ref{fig_set_up}(a). To determine the gain of the source and the beamsplitter transmissions $\eta_\text{a1}$ and $\eta_\text{b1}$, we take auxiliary measurements of laser intensities and component transmissions. 

The main issues are that $\eta_\text{a1}$ includes losses that occur inside the cell and change as a function of laser detuning, and that the gain required by the model is different from the intensity gain measured in the presence of loss. Previous studies have shown that the losses on the probe beam are much larger than those on the far detuned conjugate beam~\cite{McCormick2007}. We thus assume here that the conjugate beam does not experience any loss in the vapor. To model the losses in the source, we introduce an effective probe beam transmission, $\eta_\text{4WM}$, which contributes to $\eta_\text{a1}$ along with the ordinary optical losses, whereas $\eta_\text{b1}$ is due simply to ordinary optical losses that can be measured directly. The advantage of this model is that one is able to determine the desired $G_{\text{source}}$ and $\eta_\text{4WM}$ just by measuring the optical powers of the probe and conjugate beams with and without the presence of the pump beam. The detailed derivation is as follows.

We shine a probe beam with optical power $P_s$ into the source cell, temporarily blocking the pump beam and detuning the probe so that absorption in the atomic vapor is negligible. The probe beam experiences losses due to reflection from the cell windows, downstream optics, and the non-unity quantum efficiency of the detector. The measured optical power is

\begin{equation}
P_r = P_s\cdot {\eta^\text{off}_\text{a1}}\cdot \eta_\text{a2},
\label{Pr2}
\end{equation}
%
where $\eta^\text{off}_{\text{a1}}=0.92$ is the transmission of the source cell windows and the optical elements up to the position where the PSA cell would be inserted, and $\eta_{\text{a2}}=0.89$ is the product of the transmission of the optics after the PSA insertion point and the quantum efficiency of the detector.

With the pump beam present, and the probe frequency reset to its usual detuning, the probe power gets amplified to $P_p$ and a conjugate beam is generated with power $P_c$, where

\begin{equation}
P_p = P_s \cdot G_{\text{source}} \cdot \eta^\text{on}_\text{a1} \cdot \eta_{\text{a2}},
\label{Pp}
\end{equation}


\begin{equation}
P_c = P_s \cdot (G_{\text{source}}-1) \cdot \eta_{\text{b1}} \cdot \eta_\text{b2},
\label{Pc}
\end{equation}
and $\eta^\text{on}_{\text{a1}}=  \eta^\text{off}_\text{a1} \cdot \eta_\text{4WM}$.
Here $\eta_\text{b1}=0.94$ is the transmission of the source cell windows and the optical elements up to the position of the conjugate detector, and $\eta_\text{b2}=0.90$ is the quantum efficiency of the detector.

Solving for $P_s$ in Eq.~(\ref{Pr2}) and plugging it into Eqs.~(\ref{Pp}) and~(\ref{Pc}), one obtains

%
\begin{equation}
\frac{P_p}{P_r} = G_{\text{source}} \cdot \frac{\eta^\text{on}_\text{a1}}{\eta^\text{off}_\text{a1}}=G_{\text{source}} \cdot \eta_\text{4WM} ,
\label{Pp2}
\end{equation}

\begin{equation}
\frac{P_c}{P_r} = (G_{\text{source}}-1) \cdot \frac{\eta_\text{b1}\cdot\eta_\text{b2}}{\eta^\text{off}_\text{a1}\cdot\eta_\text{a2}}.
\label{Pc2}
\end{equation}
The gain of the source $G_{\text{source}}$ and the effective probe beam transmission $\eta_\text{4WM}$ are readily calculable from Eqs.~(\ref{Pp2}) and~(\ref{Pc2}) using the measured optical transmissions and known detector quantum efficiencies, and the measured powers $P_p$, $P_c$ and $P_r$. 

As noted in the main text, when we adjust the one photon detuning, $\Delta$, to vary the PSA gain, we also adjust the pump power in the source to maintain the same measured squeezing level.  Following this procedure we find that for $\Delta = 1.4$~GHz the effective source parameters are $G_{\text{source}}= 3.0$ and $\eta_\text{4WM}= 0.92$ and thus $\eta^\text{on}_\text{a1} = 0.85$. For $\Delta = 1.3$~GHz the effective source parameters are $G_{\text{source}}= 3.3$ and $\eta_\text{4WM}= 0.90$ and thus $\eta^\text{on}_\text{a1} = 0.83$. With these numbers plugged into the model described in Section 3, the correlation properties of the modes $a_i$ and $b_i$ can thus be fully characterized. In calculating AC noise powers one needs to account, by modifying  $\eta_\text{a2}$ and  $\eta_\text{b2}$, for the finite signal power separations from the electronic noise floors, as discussed in the main text. The intensity-difference squeezing predicted by this theory is -5.6 dB, which is very close to the experimental value of -5.8 dB. This verifies that the approximation of treating the distributed cell loss by a lumped value after the cell is sufficiently accurate for the present purposes.

\section*{Funding}
National Science Foundation (NSF) and Air Force Office of Scientific Research (AFOSR).

%

\end{document}